RESEARCH ARTICLE

# Time Scales in Epidemiological Analysis: An Empirical Comparison

Running title: Time scales in epidemiological studies

Authors:

Prabhakar Chalise, PhD*
Department of Biostatistics
University of Kansas Medical Center
Kansas City, KS 66160
Email: pchalise@kumc.edu

Eric Chicken, PhD,
Department of Statistics,
Florida State University
Tallahassee, FL, 32306
Email: chicken@stat.fsu.edu

Daniel McGee, PhD
Department of Statistics,
Florida State University
Tallahassee, FL, 32306
Email: dmcgee@fsu.edu

---

* Corresponding author
  Email: pchalise@kumc.edu
  Phone: 913-945-7987
  Fax:   913-945-7977

## Abstract

The Cox proportional hazards model is routinely used to analyze time-to-event data. To use this model requires the definition of a unique well-defined time scale. Most often, observation time is used as the time scale for both clinical and observational studies. Recently after a suggestion that it may be a more appropriate scale, chronological age has begun to appear as the time scale used in some reports. There appears to be no general consensus about which time scale is appropriate for any given analysis. It has been suggested that if the baseline hazard is exponential or if the age-at-entry is independent of covariates used in the model, then the two time scales provide similar results. In this report we provide an empirical examination of the results using the two different time scales using a large collection of data sets to examine the relationship between systolic blood pressure and coronary heart disease death (CHD death). We demonstrate, in this empirical example that the two time-scales can lead to differing results even when these two conditions appear to hold.

## 1. Introduction

Survival models are extensively used in epidemiological studies. There are various models available to analyze time to event data but the most frequently used model is the semi-parametric Cox proportional hazards (PH) model (Cox 1972). The PH model is widely used with both experimental (clinical trials) and observational data and provides a semi-parametric method of analyzing the association between a set of risk factors and the time to occurrence of an outcome. The PH model makes no assumptions concerning the nature or shape of the underlying survival distribution, but assumes a parametric form for the effect of the predictors on the hazard. In many situations, we are most interested in estimates of the model parameters since if the model is correct they provide measures of the strength of the association between a characteristic and time to event. Most statistical software contains procedures for deriving estimates of the parameters of the model.

Current literature addressing the association between a characteristic and time to event based on the analysis of observational studies using the PH model contains a mixture of analyses using the two different time scales: time-on-study and chronological age. For example, there are two widely used sets of models to predict cardiovascular disease. The models that are widely used in the United States were derived from the Framingham Heart Study (Wilson et al. 1998) using time-on-study as the time scale, while the models that are widely used in Europe (Conroy et al. 2003) were developed using age as the time scale. Similarly, there has been long-term interest in estimating the effects of obesity on mortality and two papers have appeared addressing this question but using different time scales. In 1999, Allison et al. published their estimates of the number of deaths in the U.S. that are attributable to obesity. For their analysis, they used time-on-study as the time scale. In 2007, Flegal et al. published a similar analysis, based on slightly different data, but used chronological age as the time scale. A natural question is whether at least part of the differences between the results of these analyses are due to different time scales being selected for developing the proportional hazards models used.

An extreme example was provided by Cheung et al. (2003) who demonstrated that a PH model using different time scales could result in contradictory results that the models in which the parameters have opposite sign depending on the time scale. They examined women in the Surveillance, Epidemiology, and End Results (SEER) program diagnosed with Stage I breast cancer and demonstrated that when time-on-study was used as the time scale, a younger age at diagnosis was associated with a lower mortality. If chronological age was used as the time scale, the opposite effect was found.

In this paper we provide an extensive empirical example that suggests that the use of the two different time scales with the same data can result in significantly different results and that models can disagree even if the empirical baseline hazard appears to be exponential or there is a low degree of observed correlation between the covariate and the entry-age; in some cases, although not significant, the estimated coefficients have opposite signs.

## 2. Background

In 1997, Korn et al. pointed out that the majority of published analyses based on the National Health and Nutrition Examination Survey I Epidemiological Follow up Study used time-on-study as the time scale in the PH model but suggested that chronological age might be a more appropriate time scale for some observational studies. These two scales vary in their choice of origin of the time scale. If we use chronological age as the time scale then the origin of the time scale is the date of birth. If time-on-study scale is used the origin is the date of diagnosis or date of randomization.

The mathematical formulation of the Cox model is similar for both the time-on-study and the chronological age time scales but the implicit mechanism for estimation is different (Thiebaut and Benichou 2004). In both of the formulations time is used only to order the times to event. This ordering defines the number of individuals at risk, the risk set at a particular time. Time scales that produce equivalent risk sets will produce equivalent results. The difference between the two scales is that using observation times, the risk sets are nested. That is, at any time $t$, the risk set, $R_t$ is contained in the risk set $R_{t^*}$ for any $t^* < t$. Using age does not necessarily result in this nesting of risk sets. This is because at a given age, some subjects may have already experienced failures while others may not be under study at that age and thus subjects keep on entering and exiting the study. In their paper, Korn et al. (1997) considered three possible models: (1) Age as the time scale with stratification on birth cohort, (2) Age as the time scale with no stratification and (3) Time-on-study as the time scale with baseline age as covariate. They used the National Health and Nutrition Examination Survey Epidemiological Follow-up

Study (NHEFS) data on women and compared estimated log hazard ratios using the three different models. They found no differences in log hazard ratio estimates between the first two models but reported the third one to be different from the first two. Based on this study they suggested that the most appropriate time scale might be chronological age.

Korn et al. (1997) also suggested two conditions for which the analysis results using the time-on-study time scale and the chronological age scale are equivalent. The first condition is that the baseline hazard $\lambda_0$ as a function of age is exponential, i.e., $\lambda_0(a) = c \exp(\psi a)$, where $a$ is age and $c$ and $\psi$ are constants. The second condition is that the covariate $z$ is independent of the baseline age $a_0$.

There have been two previously published examinations of the choice of time scale. Thiebaut and Benichou. (2004) and Pencina et al. (2007) conducted simulation studies to investigate the choice of time scale. Thiebaut and Benichou (2004) reported that if the cumulative hazard function is exponential, the two time scale models yielded similar regression coefficients. On the other hand they saw that the models could be significantly different even when the covariate of interest was independent of the baseline age. Pencina et al. (2007) found that when the correlation between the age-at-entry and the risk factor is zero, the biases from the two models using time-on-study time and chronological age time are very close. However, these two simulation studies are inconclusive about which time scale is the best. Also, the two papers are not consistent in their recommendations.

In this paper, we examine this issue with a large empirical datasets.

## 3. Data and Methods

### 3.1 Data

In order to examine this issue, we fitted PH models with two time-scales using a large collection of data sets from the Diverse Populations Collaboration (DPC) (McGee et al. 2005). DPC investigators pooled data from 27 observational studies to examine issues of variation in results in observational studies when differing methodologies and/or sampling methods are used for implementation and analysis. The studies can often be stratified by characteristics such as gender, area of residence (urban/rural) etc. When this stratification is accomplished there are 78 strata that we refer to as cohorts. In our analyses we excluded studies without follow-up as well as studies that did not measure blood pressure. We additionally required that all of the cohorts included in our analysis have at least 50 CHD deaths during follow-up. Our analytic data consists of information form 25 studies stratified into 54 cohorts that contained 236,624 observations among which 14,156 deaths due to Coronary Heart Disease (CHD) occurred.

### 3.2 Statistical Methods

We restrict our attention to the proportional hazards model (PH) in this paper since this is the most widely used method in epidemiological studies. Also, we focus on a single covariate, systolic blood pressure (sbp) so that we can examine whether results are consistent using the different times scales and whether they are consistent with published relationships between sbp and coronary heart disease death (CHD) as the event.

The PH model assumes that the underlying hazard (rather than survival time) is a function of the independent variables (covariates) and the contributions of the covariates to the hazard are multiplicative. This model specifies that the hazard function associated with the covariate Z satisfies,

$$\lambda(t\,|\,\mathbf{z}) = \lambda_0(t)\exp(\boldsymbol{\beta}'\mathbf{z}) \tag{1}$$

where $\lambda(t\,|\,\mathbf{z})$ is the hazard conditional on the value of the characteristics $\mathbf{z} = (z_1, z_2, ..., z_k)$, $\lambda_0(t)$ is the unspecified baseline hazard (the baseline hazard when all $z_i = 0$), $t$ is time and $\boldsymbol{\beta} = (\beta_1, \beta_2, ..., \beta_k)$ is a vector of unknown constants, the parameters of interest. Estimates of the parameters and inferences about them are based on maximum partial likelihood (Cox 1972) and the asymptotic properties are justified using martingale and counting process theory (Anderson and Gill, 1982). In the PH model, time is used to order the events and determine the risk sets of subjects still being followed when each event occurs. It is not used directly in the estimation of the coefficients of the covariates. Therefore, the different time scales in PH models lead to different estimates only if they differ in the ordering of the times to event or right censoring. Chronological age and time on study will in general produce different ordering of times to event and right censoring. Let $a_0$ be the age at which an individual enters the study, $t$ be the length of time the individual is followed until he/she experiences an event of interest or terminates participation in the study, and $a$ be the age of the individual at the point of event or censoring. We focus on the following three models in this paper:

**M1**: Time-on-study as the time scale with age at entry included as a covariate:

$$\lambda(t \mid a_0, z) = \lambda_0(t) \exp(\beta z + \gamma a_0) \qquad (2)$$

where $\beta$ is the coefficient of *z*, the *sbp* and $\gamma$ is the coefficient of $a_0$, the baseline age.

**M2**: Age as time scale without adjustment:

$$\lambda(a \mid z) = \lambda_0(a) \exp(\beta z) \qquad (3)$$

**M3**: Age as time scale with left truncation on the entry age:

$$\lambda(a \mid a_0, z) = \lambda_0(a \mid a_0) \exp(\beta z) \qquad (4)$$

where $\lambda_0(a \mid a_0)$ is the baseline hazard conditional on entry age.

We used sbp as a risk factor and fitted the three Cox PH models 2, 3 and 4 for each of the 54 cohorts separately.

There are several possible methods for comparing of the coefficients estimated from the fitted models. We used bootstrap methods to determine whether the coefficients for pairs of models differed significantly. Bootstrap samples with one thousand replications were used to calculate the standard error of the difference between two betas for each pair. The significant difference in coefficients from the two methods was calculated assuming normality of the differences:

$$\text{Test Statistics} = \frac{\hat{\beta}_1 - \hat{\beta}_2}{\text{Bootstrap SE}(\hat{\beta}_1 - \hat{\beta}_2)} \qquad (5)$$

For each of 54 cohorts, we estimated the cumulative baseline hazard function as a function of age using Breslow method (Cox 1972, with discussion). We plotted these estimates against Age to determine the shape of the baseline hazard and again we plotted the log of the estimates against the Age for the test of exponentiality (Tableman and Kim 2004). Two examples are shown in Figure 4.

We calculated correlation coefficients between sbp and age at entry to examine how strongly they were associated. This empirical correlation and whether the hazard function appeared to be exponential were used to determine if, at least approximately, the conditions from Korn et al. (1997) were met.

All analyses were performed with Stata version 9 (2005) or R version 2.15 (2008).

**4. Results**

Table 3 presents the coefficients (with their standard errors) obtained from the time-on-study time, unadjusted age time and left-truncated age time scale models. Each row of the table represents the values obtained from a separate cohort. $\beta_1$ and $se(\beta_1)$ are the coefficients and their standard errors from the time-on-study time model, $\beta_2$ and $se(\beta_2)$ are those from the unadjusted age scale model and $\beta_3$ and $se(\beta_3)$ are from the left-truncated age scale model. The p-values comparing the models, two at a time, are given in columns 8, 9 and 10. The estimated correlation coefficients between *sbp* and *age at entry* are listed in the second to last column of the table. The final column contains an indicator of whether the survival distribution for that cohort appeared to be consistent with the exponential hazard function (i.e. 1 means the shape of the cumulative baseline hazard function looks exponential and 0 means that the shape of the cumulative baseline hazard does not look exponential).

In 53 out of 54 cases, the coefficients obtained from the unadjusted age scale models are smaller than those from the time-on-study time models and in all 54 cases they are smaller than those from left-truncated age scale model. The coefficients from the time on study time model and those from the left-truncated age model are quite close. For 26 cases the coefficients from the time on study model are bigger than those from the left-truncated age model and in 27 cases they are smaller while in 1 case they are equal. Figure 1 shows the relationship between the coefficients from the time-on-study time model vs the unadjusted age time model. Similarly Figure 2 and Figure 3 present the relationships between the time-on-study time model vs left-truncated age model and unadjusted age time model vs left-truncated age model respectively. First plot in each of the figures represents the relationship between the regression coefficients from the two different models using the whole data while the remaining plots represent the same thing but subdivided by correlation categories (0.0-0.2, 0.2-0.3, 0.3-0.4, 0.4-0.5, 0.5+). It is evident from the plots that the time-on-study time model and the left-truncated age models are closer but are different from the unadjusted age time model. However, the standard errors are fairly equal in all models.

The correlation and exponentiality divide up the 54 cohorts. Table 1 presents the number of cohorts for which the estimated log hazard ratios are significantly different at the 0.05 level among the 54 cohort data sets according to the magnitude of the correlation between *sbp* and *age at entry*. All of the correlation coefficients are positive and they range from 0.0963 to 0.6280. The comparisons were made at the $\alpha = 0.05$ level of significance.

**Figure 1**: First figure represents the plot of coefficients from the time-on-study model vs unadjusted age scale model. Remaining plots represent the same plot subdivided according to the correlation categories (0.0-0.1, 0.1-0.2, 0.2-0.3, 0.3-0.4, 0.5+)

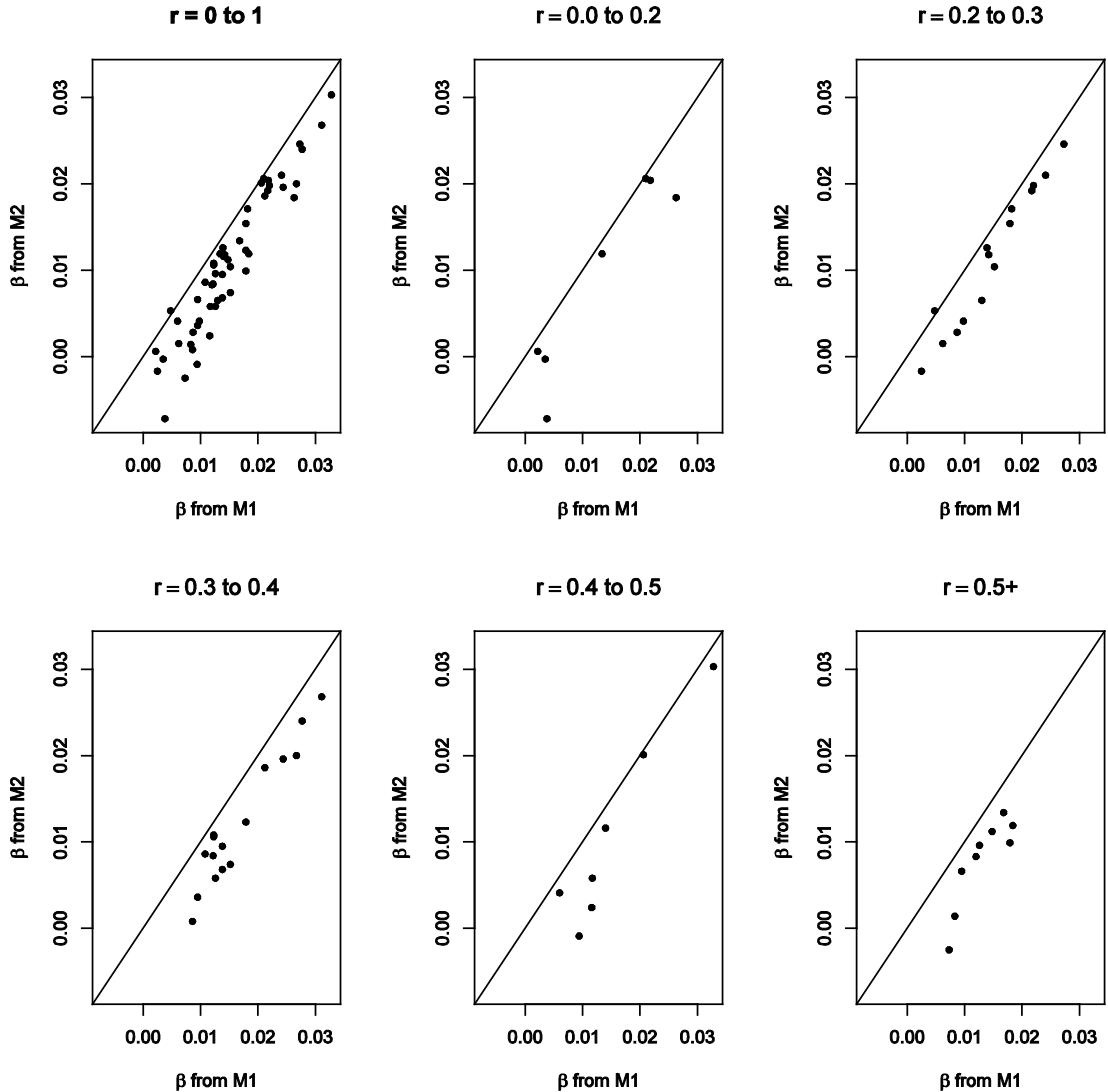

- *Time-on-study time model vs Unadjusted Age time scale model (M1 vs M2)*: The log hazard ratios estimated from models M1 and M2 are significantly different ($p<0.05$) in 40 cases out of 54 cases. When the correlation is high, above 0.5, 6 pairs of models are significantly different and 3 pairs are not different. When the value of the correlation coefficient decreases, still many significant cases appear. When the correlation is modestly strong (0.3-0.4), 12 cases are significant and 4 cases are non-significant. Both Korn (1997) and Pencina (2007) suggest that we should expect the models to be equivalent at independence or correlation zero. But, their argument does not extend to low or moderate correlations. At correlations between 0.1 and 0.2 (second last row of the table) the models are significantly different in all 6 cases. These results support that the models can perform differently even when there is very little association between the risk factor and the baseline age.

**Figure 2**: First figure represents the plot of coefficients from the time-on-study model vs left truncated age scale model. Remaining plots represent the same plot subdivided according to the correlation categories (0.0-0.1, 0.1-0.2, 0.2-0.3, 0.3-0.4, 0.5+)

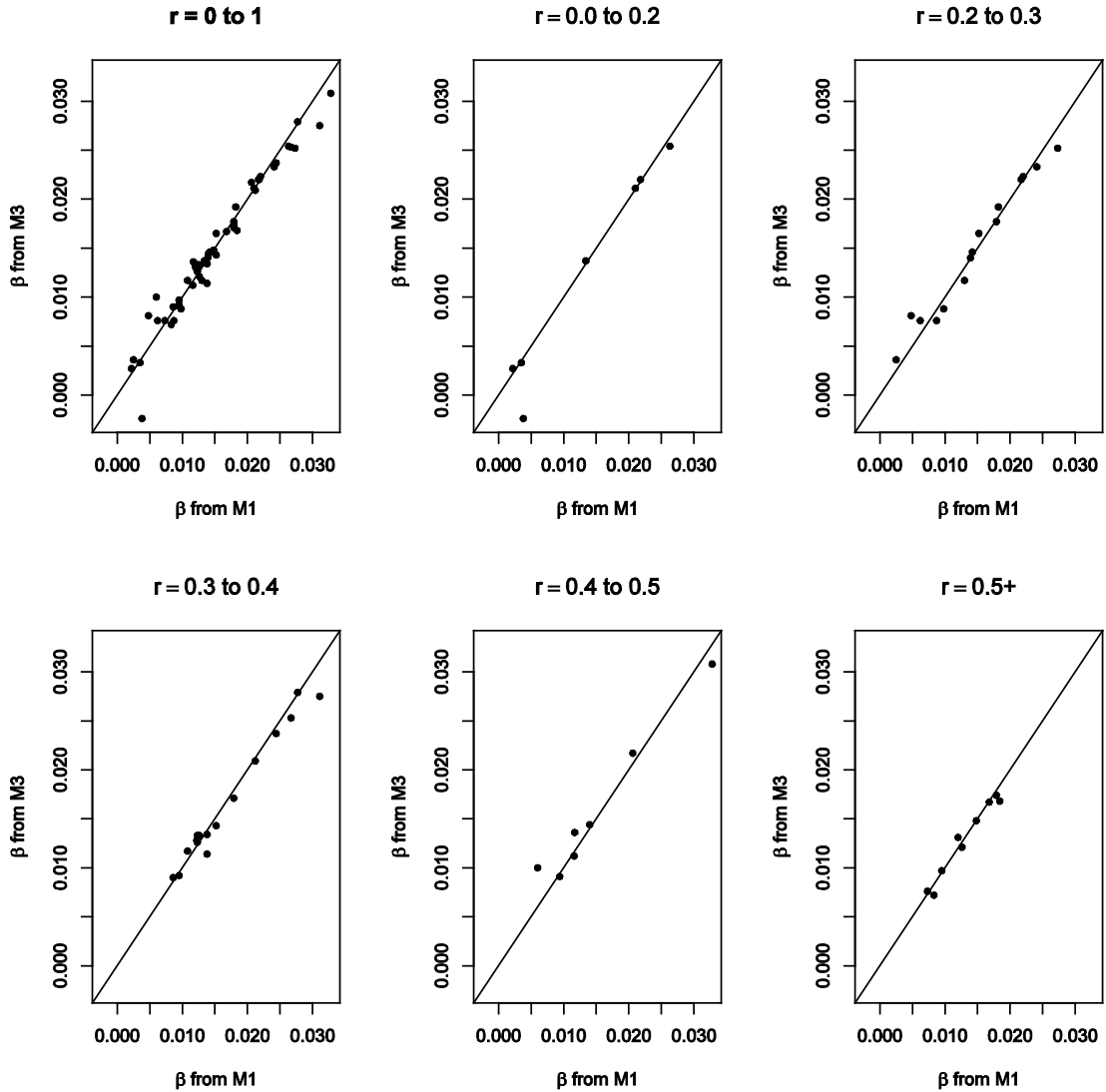

- *Time-on-study time model vs Left-truncated age time scale model (M1 vs M3)*: The results from models M1 and M3 are similar in 51 cases and are different in 3 cases. When the correlation is above 0.4 and below 0.2 none of the pairs of the models differ significantly but some cases are significant at the moderate values of the correlation (0.2 to 0.4). 1 case is significant out of 16 when correlation is between 0.3 and 0.4 and 2 cases are significant out of 15 when the correlation is in the range of 0.2 to 0.3. This shows no obvious influence of the correlation on whether the log hazards ratio estimates are significantly different between models. These are the closest pair of models in terms of predictor significance out of three pairs of comparisons we made.

**Figure 3**: First figure represents the plot of coefficients from the left truncated age scale model vs unadjusted age scale model. Remaining plots represent the same plot subdivided according to the correlation categories (0.0-0.1, 0.1-0.2, 0.2-0.3, 0.3-0.4, 0.5+)

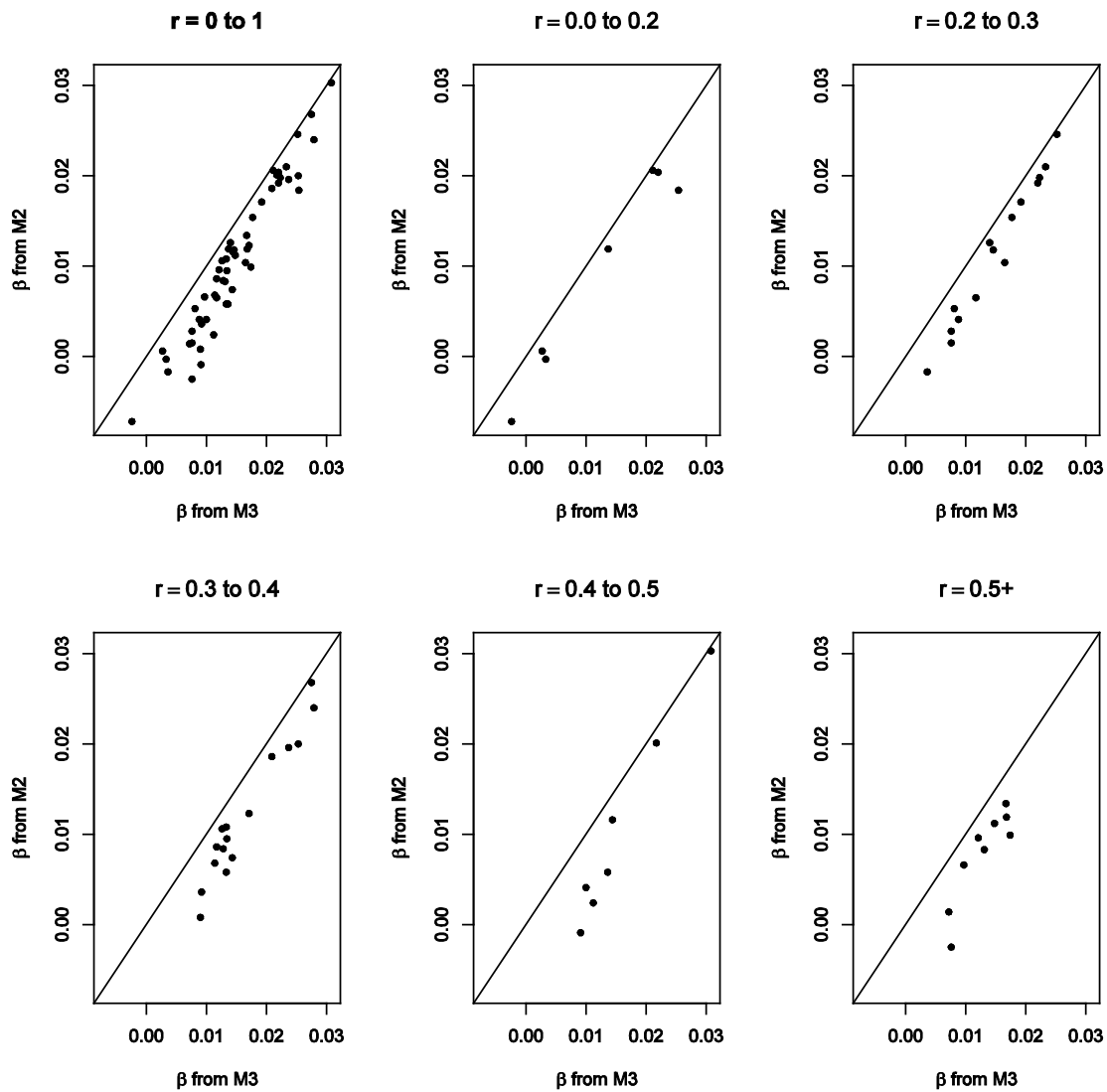

- *Unadjusted age time model vs Left-truncated age time scale model (M2 vs M3)*: These two models have the most extensive differences. The estimated log hazard ratios are significantly different between the unadjusted age time model and the left-truncated age time model in 53 cases. No matter how strong or how weak the association between sbp with baseline age these pairs of models are always different except in one case. At correlation coefficient between 0.4 and 0.5, 1 case is non-significant.

**Table1**: Table showing the number of cohorts for which the estimated log hazard ratios are significantly different between models by correlation between the SBP and age at entry. Here, Left trunc stands for left truncated age time scale.

| Correlation | Time-on-study vs Unadj. Age | | Time-on-study vs Left trunk Age | | Unadj. Age vs Left trunk Age | |
|---|---|---|---|---|---|---|
| | Sig | Non-sig | Sig | Non-sig | Sig | Non-sig |
| 0.5+ | 6 | 3 | 0 | 9 | 9 | 0 |
| 0.4-0.5 | 4 | 3 | 0 | 7 | 6 | 1 |
| 0.3-0.4 | 12 | 4 | 1 | 15 | 16 | 0 |
| 0.2-0.3 | 12 | 3 | 2 | 13 | 15 | 0 |
| 0.1-0.2 | 6 | 0 | 0 | 6 | 6 | 0 |
| 0.0-0.1 | 0 | 1 | 0 | 1 | 1 | 0 |
| Total | 40 | 14 | 3 | 51 | 53 | 1 |

**Figure 4:** Two examples of cumulative baseline hazard and corresponding log-cumulative hazard plots. Example 1 represents an example of exponential case and Example 2 represents an example of non-exponential case.

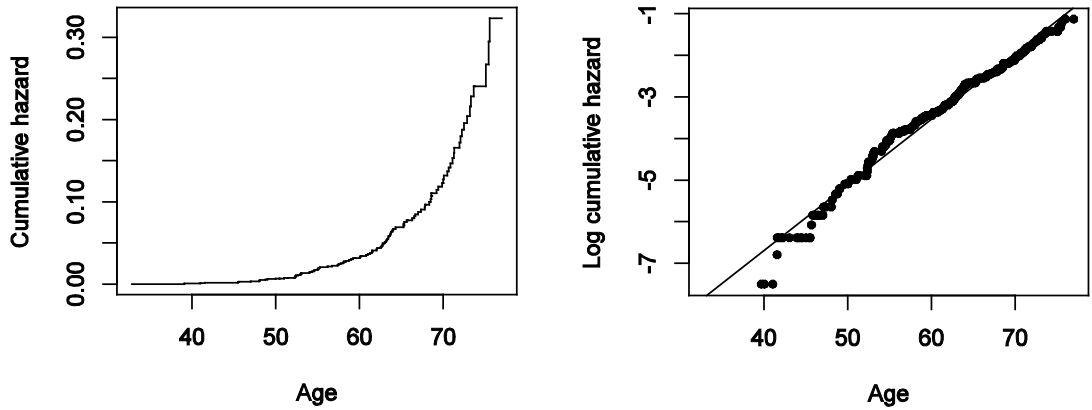

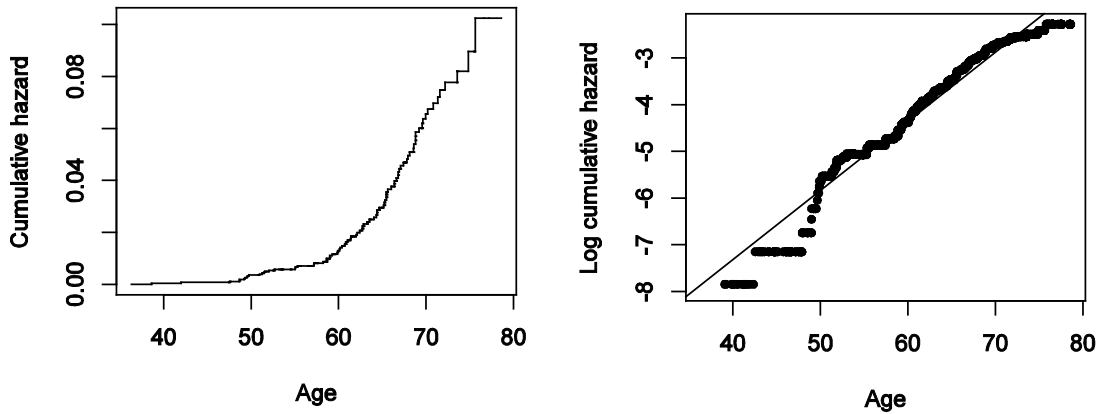

Next, we estimated the cumulative baseline hazard function as a function of age using Breslow's method (1972) and plotted them against age. We further plotted the log of the cumulative hazard estimate against age to see if it appears linear. The results show that 18 cases appear close to the exponential form hazard and the remaining 36 plots look non-exponential form.

Table 2 includes the number of cases showing whether the results from three pairs of models are different or not according to the shape of the cumulative baseline hazard function. The log hazard estimates from the time-on-study time and unadjusted age time scale models are significantly different in 14 cases even though the cumulative baseline hazard functions appear exponential, and only in 4 cases do the two results agree. Similarly, time-on-study time and left-truncated age time models are different in 1 case with the baseline hazard exponential. Likewise, unadjusted age and left truncated age scale models do not agree in all 18 cases.

**Table 2:** Table showing the number of cohorts for which the estimated log hazard ratios are significantly different between models by whether the baseline hazard appeared to be exponential.

|  | Time-on-study vs Unadj. Age | | Time-on-study vs Left trunk Age | | Unadj. Age vs Left trunk Age | |
| --- | --- | --- | --- | --- | --- | --- |
| Exponential | Sig | Non-sig | Sig | Non-sig | Sig | Non-sig |
| Yes | 14 | 4 | 1 | 17 | 18 | 0 |
| No | 26 | 10 | 2 | 34 | 35 | 1 |
| Total | 40 | 14 | 3 | 51 | 53 | 1 |

We also noted that in some of the cohort data sets the time-on-study time and chronological age time scale models indicate the association between the risk factor and hazard of occurring disease are in opposite direction. In five cohort data sets, the coefficients from unadjusted age time scale models and in one case the coefficient from left truncated age time scale model are negative. These results agree with those reported by Cheung et al. (2003). The widely accepted fact is that the systolic blood pressure increases risk of coronary heart disease deaths. But, we observed that age scale models can not always detect this, instead sometimes it erroneously suggests that sbp has beneficial effect on the coronary heart disease.

(Table 3 is at the end of the paper)

## 5. Discussion

In this report we presented a comparison of results of PH models using two different time scales in 54 different datasets. In our results, we found that using unadjusted age as the time scale results in significantly different coefficients even when there is very low correlation between a covariate and baseline age and also when the cumulative baseline hazards appear to be exponentially distributed. We also found that the estimates when unadjusted age was used as the time scale are almost universally lower than the estimates that result from using the other time scales. However, the estimates obtained from the time-on-study and left truncated age time scale models are closer in most of the cases.

Our empirical example coincides with what one would expect given the method of estimation for the PH model. To estimate the coefficients we use the partial likelihood function in which the baseline hazard is factored out and the time scale provides the method of ordering the times to event and determining risk sets (Chalise et al. (2012), Thiebaut et al. (2004)). Pencina et al. (2007) notes that if all ages at baseline were the same, the models would all be equivalent. The distribution of ages at baseline plays a central role in determining when the time scales will produce equivalent results. Chalise et al. have shown how the baseline age influences the estimation of the parameters in their paper (Chalise et al. 2012).

It is extremely valuable to use real datasets to assess the different models in empirical analysis as presented in this paper. A shortcoming of doing so is that the comparisons are based on some data generating mechanism we have no clue about. Chalise et al. (2013) have carried out extensive simulation studies to address the question of robustness using one time scale when the other is actually the correct one (Chalise et al. 2013). They generated data according to a specified model and then compared the different models against the specified model with respect to bias, mean square error and measure of predictive discrimination. Two simulated scenarios were created where they correctly specified one of the two time scales. When time-on-study was correctly specified, the time on study models were better with respect to all three measures. But, when age was the correct time scale both time-scale models performed approximately equally well. This simulation studies suggested that the time-on-study models are robust to misspecification of the underlying time scale suggesting that time-on-study models may be preferable in case of uncertainty of the true time scale.

In some situations, there may be multiple plausible time scales. For example, automobile warranties usually use two time scales, calendar time and cumulative mileage; in studies of skin cancer among occupationally exposed workers, cumulative exposure to radiation may provide a better time scale than does the person's age or time on study. Some investigators have examined methods for deriving an optimal time scale.

Farewell and Cox (1979) and Oakes (1995) suggested choosing a time scale that combines two or more times scales in such a way that the resulting scale accounts for as much the variation as possible. Duchesne and Lawless (2000) introduced the concept of an ideal time scale. Their work, however, focuses on usage (e.g. mileage) or exposure (e.g. asbestos exposure) variables that used as time scales (and could be adapted to epidemiological analyses in some instances) but do not solve the problems inherent in comparing time scales differ only in having different origins.

A unique well-defined time scale is indispensable for event history analysis. Given the lack of an agreed upon definition for an optimal or even correct time scale, robustness may be the only practical criteria on which to base our decision. In the meantime, our personal opinion, based on our results is that the time-on-study time scale is usually appropriate since it answers the conditional question that is the primary focus of epidemiological studies: Given what we measure at baseline what is the probability of future events?

**Conflict of Interest:** The authors have declared no conflict of interest.

**Table 3**: Coefficients, Standard Errors, Pairwise p-values and Correlations

| | Time-on-Study [1] | | Unadj. Age [2] | | Left Trunc Age [3] | | p-values | | | | | | | |
|---|---|---|---|---|---|---|---|---|---|---|---|---|---|---|
| Sl | beta1 | se1 | beta2 | se2 | beta3 | se3 | [1] and [2] | | [1] and [3] | | [2] and [3] | | Corr | Expo |
| 1 | 0.0062 | 0.0065 | 0.0015 | 0.0066 | 0.0076 | 0.0067 | 0.2162 | | 0.6974 | | 0.0000 | * | 0.2369 | 1 |
| 2 | 0.0022 | 0.0035 | 0.0006 | 0.0035 | 0.0027 | 0.0034 | 0.0014 | * | 0.2113 | | 0.0027 | * | 0.1321 | 0 |
| 3 | 0.0025 | 0.0019 | -0.0017 | 0.0019 | 0.0036 | 0.0019 | 0.0000 | * | 0.0000 | * | 0.0000 | * | 0.2066 | 0 |
| 4 | 0.0120 | 0.0039 | 0.0083 | 0.0034 | 0.0131 | 0.0035 | 0.2020 | | 0.6722 | | 0.0006 | * | 0.5027 | 1 |
| 5 | 0.0138 | 0.0046 | 0.0068 | 0.0040 | 0.0114 | 0.0042 | 0.0727 | | 0.4388 | | 0.0365 | * | 0.3717 | 0 |
| 6 | 0.0184 | 0.0038 | 0.0119 | 0.0033 | 0.0168 | 0.0034 | 0.0489 | * | 0.6058 | | 0.0011 | * | 0.5785 | 1 |
| 7 | 0.0060 | 0.0036 | 0.0041 | 0.0031 | 0.0100 | 0.0034 | 0.4473 | | 0.0568 | | 0.0002 | * | 0.4359 | 0 |
| 8 | 0.0087 | 0.0051 | 0.0028 | 0.0050 | 0.0076 | 0.0051 | 0.0000 | * | 0.3974 | | 0.0000 | * | 0.2528 | 0 |
| 9 | 0.0035 | 0.0038 | -0.0003 | 0.0037 | 0.0033 | 0.0038 | 0.0113 | * | 0.8026 | | 0.0011 | * | 0.1455 | 0 |
| 10 | 0.0267 | 0.0045 | 0.0200 | 0.0045 | 0.0253 | 0.0045 | 0.0008 | * | 0.3507 | | 0.0009 | * | 0.3933 | 0 |
| 11 | 0.0123 | 0.0027 | 0.0106 | 0.0026 | 0.0126 | 0.0026 | 0.2880 | | 0.8303 | | 0.0124 | * | 0.3981 | 1 |
| 12 | 0.0123 | 0.0034 | 0.0108 | 0.0033 | 0.0133 | 0.0032 | 0.5922 | | 0.6892 | | 0.0372 | * | 0.3681 | 0 |
| 13 | 0.0168 | 0.0028 | 0.0134 | 0.0026 | 0.0167 | 0.0026 | 0.0589 | | 0.9468 | | 0.0002 | * | 0.5358 | 1 |
| 14 | 0.0179 | 0.0025 | 0.0123 | 0.0023 | 0.0171 | 0.0023 | 0.0077 | * | 0.6171 | | 0.0001 | * | 0.3939 | 0 |
| 15 | 0.0206 | 0.0024 | 0.0201 | 0.0021 | 0.0217 | 0.0021 | 0.7389 | | 0.4320 | | 0.0001 | * | 0.4704 | 0 |
| 16 | 0.0217 | 0.0022 | 0.0192 | 0.0021 | 0.0220 | 0.0021 | 0.0124 | * | 0.7389 | | 0.0000 | * | 0.2325 | 0 |
| 17 | 0.0048 | 0.0058 | 0.0053 | 0.0057 | 0.0081 | 0.0058 | 0.8026 | | 0.0989 | | 0.0001 | * | 0.2633 | 0 |
| 18 | 0.0073 | 0.0047 | -0.0025 | 0.0044 | 0.0076 | 0.0049 | 0.0001 | * | 0.8745 | | 0.0000 | * | 0.6280 | 0 |
| 19 | 0.0094 | 0.0034 | -0.0009 | 0.0033 | 0.0091 | 0.0035 | 0.0000 | * | 0.8303 | | 0.0000 | * | 0.4660 | 0 |
| 20 | 0.0328 | 0.0048 | 0.0303 | 0.0045 | 0.0308 | 0.0045 | 0.3173 | | 0.4047 | | 0.2113 | | 0.4964 | 0 |
| 21 | 0.0311 | 0.0037 | 0.0268 | 0.0036 | 0.0275 | 0.0036 | 0.0114 | * | 0.0244 | * | 0.0196 | * | 0.3558 | 0 |
| 22 | 0.0126 | 0.0043 | 0.0058 | 0.0042 | 0.0133 | 0.0043 | 0.0001 | * | 0.5597 | | 0.0000 | * | 0.3383 | 0 |
| 23 | 0.0152 | 0.0049 | 0.0074 | 0.0049 | 0.0143 | 0.0049 | 0.0000 | * | 0.4132 | | 0.0000 | * | 0.3896 | 1 |
| 24 | 0.0095 | 0.0053 | 0.0036 | 0.0052 | 0.0092 | 0.0053 | 0.0002 | * | 0.8026 | | 0.0010 | * | 0.3312 | 0 |
| 25 | 0.0116 | 0.0046 | 0.0024 | 0.0045 | 0.0112 | 0.0045 | 0.0000 | * | 0.7389 | | 0.0000 | * | 0.4168 | 1 |
| 26 | 0.0218 | 0.0020 | 0.0204 | 0.0019 | 0.0220 | 0.0020 | 0.0196 | * | 0.6892 | | 0.0000 | * | 0.1917 | 0 |
| 27 | 0.0179 | 0.0021 | 0.0154 | 0.0021 | 0.0177 | 0.0021 | 0.0000 | * | 0.6171 | | 0.0000 | * | 0.2985 | 1 |
| 28 | 0.0142 | 0.0014 | 0.0118 | 0.0014 | 0.0146 | 0.0014 | 0.0000 | * | 0.3173 | | 0.0000 | * | 0.2186 | 0 |
| 29 | 0.0212 | 0.0013 | 0.0186 | 0.0012 | 0.0209 | 0.0012 | 0.0000 | * | 0.5485 | | 0.0000 | * | 0.3217 | 1 |
| 30 | 0.0179 | 0.0053 | 0.0099 | 0.0051 | 0.0174 | 0.0052 | 0.0014 | * | 0.8118 | | 0.0007 | * | 0.5272 | 0 |
| 31 | 0.0086 | 0.0044 | 0.0008 | 0.0043 | 0.0090 | 0.0044 | 0.0000 | * | 0.8242 | | 0.0000 | * | 0.3849 | 1 |
| 32 | 0.0083 | 0.0066 | 0.0014 | 0.0063 | 0.0072 | 0.0065 | 0.0553 | | 0.6722 | | 0.0203 | * | 0.5318 | 0 |
| 33 | 0.0117 | 0.0047 | 0.0058 | 0.0047 | 0.0136 | 0.0047 | 0.0032 | * | 0.2912 | | 0.0000 | * | 0.4051 | 1 |
| 34 | 0.0263 | 0.0082 | 0.0184 | 0.0079 | 0.0254 | 0.0081 | 0.0024 | * | 0.6357 | | 0.0000 | * | 0.1744 | 0 |
| 35 | 0.0038 | 0.0093 | -0.0072 | 0.0089 | -0.0024 | 0.0091 | 0.0145 | * | 0.1777 | | 0.0002 | * | 0.1904 | 0 |
| 36 | 0.0098 | 0.0045 | 0.0041 | 0.0044 | 0.0088 | 0.0045 | 0.0000 | * | 0.2113 | | 0.0000 | * | 0.2293 | 0 |
| 37 | 0.0130 | 0.0042 | 0.0065 | 0.0041 | 0.0117 | 0.0042 | 0.0000 | * | 0.1042 | | 0.0000 | * | 0.2487 | 0 |
| 38 | 0.0095 | 0.0035 | 0.0066 | 0.0035 | 0.0097 | 0.0034 | 0.0383 | * | 0.8557 | | 0.0019 | * | 0.5518 | 1 |
| 39 | 0.0138 | 0.0034 | 0.0095 | 0.0034 | 0.0134 | 0.0034 | 0.0114 | * | 0.7389 | | 0.0053 | * | 0.3837 | 1 |
| 40 | 0.0126 | 0.0018 | 0.0096 | 0.0018 | 0.0121 | 0.0018 | 0.0000 | * | 0.3173 | | 0.0000 | * | 0.5763 | 0 |
| 41 | 0.0122 | 0.0017 | 0.0084 | 0.0016 | 0.0128 | 0.0016 | 0.0000 | * | 0.2301 | | 0.0000 | * | 0.3994 | 0 |
| 42 | 0.0140 | 0.0029 | 0.0116 | 0.0028 | 0.0144 | 0.0029 | 0.0077 | * | 0.5677 | | 0.0000 | * | 0.4904 | 1 |
| 43 | 0.0108 | 0.0023 | 0.0086 | 0.0023 | 0.0117 | 0.0023 | 0.0002 | * | 0.0719 | | 0.0000 | * | 0.3183 | 1 |
| 44 | 0.0273 | 0.0032 | 0.0246 | 0.0032 | 0.0252 | 0.0032 | 0.0027 | * | 0.0196 | * | 0.0027 | * | 0.2412 | 1 |
| 45 | 0.0210 | 0.0017 | 0.0206 | 0.0017 | 0.0211 | 0.0017 | 0.1824 | | 0.6171 | | 0.0000 | * | 0.0963 | 0 |
| 46 | 0.0241 | 0.0037 | 0.0210 | 0.0035 | 0.0233 | 0.0036 | 0.0268 | * | 0.4670 | | 0.0010 | * | 0.2260 | 0 |

| 47 | 0.0220 | 0.0021 | 0.0198 | 0.0020 | 0.0223 | 0.0020 | 0.0002 | * | 0.5485 | 0.0000 | * | 0.2393 | 0 |
| 48 | 0.0139 | 0.0015 | 0.0126 | 0.0015 | 0.0140 | 0.0015 | 0.0012 | * | 0.8026 | 0.0000 | * | 0.2771 | 0 |
| 49 | 0.0134 | 0.0013 | 0.0119 | 0.0012 | 0.0137 | 0.0013 | 0.0000 | * | 0.3173 | 0.0000 | * | 0.1978 | 0 |
| 50 | 0.0182 | 0.0018 | 0.0171 | 0.0017 | 0.0192 | 0.0017 | 0.0668 |   | 0.0956 | 0.0000 | * | 0.2689 | 0 |
| 51 | 0.0148 | 0.0028 | 0.0112 | 0.0027 | 0.0148 | 0.0028 | 0.0455 | * | 1.0000 | 0.0056 | * | 0.5532 | 1 |
| 52 | 0.0152 | 0.0027 | 0.0104 | 0.0025 | 0.0165 | 0.0027 | 0.0115 | * | 0.2787 | 0.0007 | * | 0.2619 | 1 |
| 53 | 0.0277 | 0.0048 | 0.0240 | 0.0047 | 0.0279 | 0.0047 | 0.0515 |   | 0.8940 | 0.0004 | * | 0.3510 | 0 |
| 54 | 0.0244 | 0.0034 | 0.0196 | 0.0032 | 0.0237 | 0.0033 | 0.0027 | * | 0.6171 | 0.0000 | * | 0.3933 | 0 |
|   | | | | | 1- Exponential, 0-Non-exponential | | | | | | | | 18 |